\documentclass[prd,preprintnumbers,superscriptaddress,nofootinbib,twocolumn,aps,10pt]{revtex4-2}
\usepackage{epsfig}
\usepackage{amsmath}
\usepackage{amsfonts}
\usepackage{float}
\usepackage{amssymb}
\usepackage{slashed}
\usepackage[dvipsnames]{xcolor}
\usepackage{pbox}
\usepackage{subfigure}
\usepackage[colorlinks,citecolor=blue, linkcolor = blue, urlcolor = blue]{hyperref}
\usepackage{tabularx}
\usepackage{titlesec}
\usepackage{scrextend}
\usepackage{enumitem}
\usepackage[normalem]{ulem}
\usepackage{orcidlink}
\usepackage{natbib}

\def\gagg{g_{a \gamma \gamma}}
\def\gae{g_{aee}}
\def\me{m_e}
\def\Lag{\mathcal{L}}
\def\bal#1\eal{\begin{align}#1\end{align}}
\def\bea#1\eea{\begin{eqnarray}#1\end{eqnarray}}
\def\beq#1\eeq{\begin{equation}#1\end{equation}}
\def\del{\partial}
\def\nn{\nonumber}

\begin{document}


\title{Limits on the axion–photon coupling from Chandrayaan-2 observations}

\author{Tanmoy Kumar\orcidlink{0000-0001-9775-6645}}
 \email{kumartanmoy1998@gmail.com}
 \affiliation{School of Physical Sciences, Indian Association for the Cultivation of Science, 2A \& 2B Raja S.C. Mullick Road, Jadavpur, Kolkata 700032, India}

\author{N. P. S. Mithun\orcidlink{0000-0003-3431-6110}}%
 \email{mithun@prl.res.in}
  \affiliation{Physical Research Laboratory, Navrangpura, Ahmedabad 380009, India}

\author{Subhendra Mohanty\orcidlink{0000-0003-0070-6647}}
 \email{subhendram@iiserb.ac.in}
 \affiliation{Department of Physics, Indian Institute of Technology Kanpur, Kanpur 208016, India}
  \affiliation{Department of Physics, Indian Institute of Science Education and Research Bhopal,\\
Bhopal 462066, India}

\author{Sourov Roy\orcidlink{0000-0002-1015-3241}}%
 \email{tpsr@iacs.res.in}
  \affiliation{School of Physical Sciences, Indian Association for the Cultivation of Science, 2A \& 2B Raja S.C. Mullick Road, Jadavpur, Kolkata 700032, India}

\author{B.S. Bharath Saiguhan\orcidlink{0000-0001-7580-364X}}%
 \email{bsg@prl.res.in}
  \affiliation{Physical Research Laboratory, Navrangpura, Ahmedabad 380009, India}

  \author{Santosh Vadawale\orcidlink{0000-0002-2050-0913}}%
 \email{santoshv@prl.res.in}
  \affiliation{Physical Research Laboratory, Navrangpura, Ahmedabad 380009, India}

\begin{abstract}
Axions and axion-like particles (ALPs) have gained immense attention in searches for beyond Standard Model (BSM) physics. Experiments searching for axions leverage their predicted couplings to Standard Model (SM) particles to look for observable signals. Though weak, these couplings allow axions to be produced abundantly in the interiors of stars such as the Sun. Once created, axions can escape the Sun and while passing through the solar atmosphere, oscillate into photons in the magnetic field producing x-rays. For the first time, we used data from the observation of soft x-rays from the quiet Sun during the 2019-20 solar minimum by the solar x-ray monitor (XSM), onboard India’s Chandrayaan-2 lunar exploration mission, to constrain the coupling of axions to photons ($\gagg$). Using the latest models of the solar atmosphere to calculate the magnetic field and plasma frequency, we constrain \textcolor{black}{$\gagg \lesssim (0.50\,-\,2.26) \times 10^{-10}$ GeV$^{-1}$} at 95\% confidence level for axion masses $m_a \lesssim 5 \times 10^{-4}$eV.

\end{abstract}

\maketitle


\textit{Introduction.} Axions and axion-like particles (ALPs) are well-motivated candidates for physics beyond the Standard Model (SM). They arise generically as pseudo-Nambu--Goldstone bosons (pNGBs) of spontaneously broken approximate global $U(1)$ symmetries in many extensions of the SM~\cite{Kim:2008hd, DiLuzio:2020wdo, Arvanitaki:2009fg}. The best-known example is the QCD axion, originating from the Peccei--Quinn (PQ) solution to the strong CP problem~\cite{Peccei:1977hh,Peccei:1977ur,Weinberg:1977ma,Wilczek:1977pj,Peccei:2006as}. Throughout this work, we use the term axions to denote generic pNGBs associated with nonlinearly realized approximate global symmetries.

Axions are predicted to have masses ranging, from $10^{-22}\,\mathrm{eV}$ to several GeV, and interact extremely weakly with SM particles. Consequently, light axions ($m_a \lesssim \mathrm{few}\,\mathrm{eV}$) are compelling cold dark matter (DM) candidates~\cite{Preskill:1982cy,Abbott:1982af,Dine:1982ah,Khlopov:1999tm}. Hence, most experimental searches for light axions target axion DM through its coupling to photons~\cite{Ouellet:2018beu,Ouellet:2019tlz,Silva-Feaver:2016qhh,DMRadio:2022jfv,DMRadio:2022pkf,ADMX:2003rdr,ADMX:2018gho,ADMX:2019uok,HAYSTAC:2018rwy,Caldwell:2016dcw,Millar:2016cjp}. However, axions need not constitute the DM, motivating complementary search strategies.

Astrophysical objects such as stars provide powerful laboratories to search for axions. Owing to their high temperatures, densities, and volumes, they can efficiently produce axions~\cite{Vysotsky:1978dc,Raffelt:2006cw,Lella:2022uwi,Lella:2023bfb,Carenza:2024ehj,Lecce:2025dbz,Fiorillo:2025gnd}, which can freely stream out of the stars for sufficiently small couplings. In particular, the Sun is a prime target: axions produced in its core at $T\sim\mathrm{few}\,\mathrm{keV}$ and $\rho\sim150\,\mathrm{g\,cm^{-3}}$ escape with $\mathcal{O}($keV$)$ energies. The CERN Axion Solar Telescope (CAST) has searched for such solar axions and, in the absence of a signal, has placed stringent bounds on axion parameter space~\cite{CAST:2004gzq,Barth:2013sma,CAST:2017uph}.

In this work, we explore a novel avenue for detecting solar axions by using the solar soft x-ray spectrum measured by the solar x-ray Monitor (XSM) on board the Chandrayaan-2 lunar orbiter of the Indian Space Research Organization (ISRO), during the 2019-2020 solar minimum~\cite{2021ApJ...912L..12V,Vadawale:2021pis}. This soft x-ray spectrum consists of two components - the background quiet-Sun emission and the emission from the flaring plasma. Additionally, solar axions will also contribute to the total solar soft x-ray emission due to their conversion into photons in the solar magnetic field. We conduct a search for this axion-induced x-ray emission from the Sun using the observed solar soft x-ray spectrum and put constraints on the axion-photon coupling. We find no evidence of axion-induced signal over the background and consequently set an upper bound on the axion-photon coupling, which is comparable to the present constraints from the CAST experiment and is limited only by the collecting area of the XSM.

A similar analysis using data from the NuSTAR observation of the Sun during the solar minimum has been presented in~\cite{Ruz:2024gkl}. However, unlike NuSTAR, which was able to observe only a portion of the Sun because of its small field of view (FOV) ($12' \times 12'$), the XSM
observes emission from the entire solar disk.
Our work provides a compelling case for having a dedicated solar observatory operational during the next solar minimum which can significantly improve upon these bounds in an independent fashion.

\vspace*{0.5cm}

\textit{Axion production inside the Sun.} We start with the following Lagrangian for the axion field $a$:
\bal
\Lag \supset\, &\dfrac{1}{2}\, \del_\mu a\, \del^\mu a\, -\,  \dfrac{1}{2}\, m_a^2\,a^2\,\nn\\ &+\,\dfrac{\gae}{2 \me}\, \del_\mu a\, \Bar{\psi}_e\gamma^\mu\gamma_5 \psi_e\, -\, \dfrac{\gagg}{4}\, a F^{\mu \nu} \Tilde{F}_{\mu \nu},
\label{eqn:lagrangian}
\eal
where $\gagg$ is the axion-photon coupling, $\gae$ is the axion-electron coupling, $m_a$ is the mass of the axion, $\psi_e$ is the electron field and $F^{\mu \nu}$ is the electromagnetic field strength tensor. For simplicity we do not consider any particular axion model and treat $\gagg$ as an independent parameter alongside $m_a$.

In the solar plasma where there exists huge quantities of free electrons and ions, due to the above couplings axions can be produced abundantly via the following processes:
\begin{enumerate}[itemsep=0.1pt,topsep=0.2pt]
    \item Primakoff process: $\gamma\, +\, (e^-, Ze) \to (e^-, Ze)\, +\, a$
    \item Compton scattering: $\gamma\, +\, e \to e\, +\, a$
    \item e-ion bremsstrahlung: $e\, +\, Ze \to e\, +\, Ze\, +\, a$
    \item e-e bremsstrahlung: $e\, +\, e \to e\, +\, e\, +\, a$
\end{enumerate}
where $e$ stands for electron and $Ze$ stands for a charged ion. The amplitude of the Primakoff process is governed by $\gagg$ while all that of all the other processes are governed by $\gae$.

Following~\cite{Raffelt:1985nk} we calculate the axion emission rate per unit volume and we integrate it over a standard solar model~\cite{Bahcall:2004pz} and finally dividing by the total surface area of the Sun, we obtain the flux of axions coming out of the Sun. For the light axions ($m_a << \rm eV$) that we shall consider in this work, the spectral axion fluxes (in units of cm$^{-2}$ s$^{-1}$ keV$^{-1}$) produced by the Sun via the four processes are well approximated by the following fitting formulae~\cite{Barth:2013sma}
\bal
\left. \dfrac{d \phi_a}{d E_a} \right \vert_P &= 3.0 \times 10^{11}\, \left( \dfrac{\gagg}{10^{-12} \text{GeV}^{-1}} \right)^2 E_a^{2.450}\, e^{-0.829 E_a},\,\nn \\
\left. \dfrac{d \phi_a}{d E_a} \right \vert_B &= 1.25 \times 10^{14}\, \left( \dfrac{\gae}{10^{-13}} \right)^2 \dfrac{E_a}{1 + 0.667\, E_a^{1.278}}\, e^{-0.77 E_a},\, \nn \\
\left. \dfrac{d \phi_a}{d E_a} \right \vert_C &= 6.32 \times 10^{11}\, \left( \dfrac{\gae}{10^{-13}} \right)^2 E_a^{2.987}\, e^{-0.776 E_a},
\label{eq:solar_axion_flux_PBC}
\eal
where $E_a$ is the axion energy in keV and $P,\, B,\, \text{and}\, C$ stand for Primakoff, Bremsstrahlung and Compton respectively. In order to calculate the bremsstrahlung flux, both the electron-electron bremsstrahlung and electron-ion bremsstrahlung have been taken into account.

\begin{figure}[ht!]
    \centering
    \includegraphics[width=0.8\linewidth]{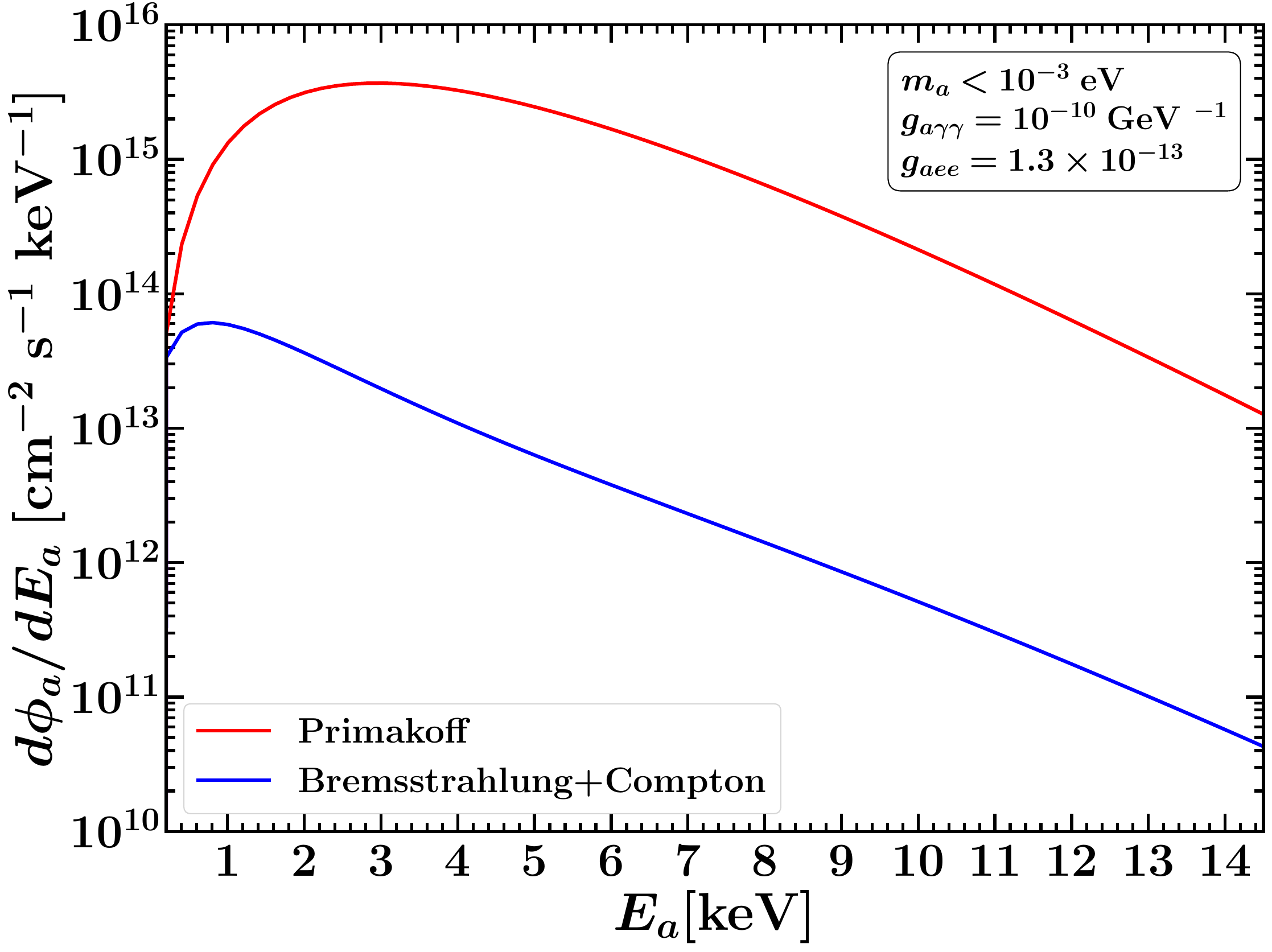}
    \caption{The solar axion flux at the solar surface arising from the Primakoff process (red solid) and from the combined Bremsstrahlung and Compton processes (blue solid). The couplings are set to $\gagg = 10^{-10}$ GeV$^{-1}$ and $\gae = 1.3 \times 10^{-13}$.}
    \label{fig:axion_flux}
\end{figure}

Fig.~\ref{fig:axion_flux} shows the fluxes of solar axions at the solar surface produced via the Primakoff process (solid red line) and through the combined Bremsstrahlung and Compton processes (solid blue line), for \textcolor{black}{$\gagg = 10^{-10}\,\text{GeV}^{-1}$} and $\gae = 1.3\times10^{-13}$. The total solar axion flux is obtained as the sum of the individual contributions from the Primakoff, Bremsstrahlung, and Compton processes, and therefore depends on both $\gagg$ and $\gae$. Within the parameter space and energy range considered here, the Primakoff-induced flux is substantially larger than the combined Bremsstrahlung and Compton components. Nevertheless, for completeness, all production channels are included in the subsequent analysis.

\vspace*{0.5cm}
\textit{Axion-photon conversion outside the Sun.} Relativistic axions produced in the solar core propagate along nearly radial trajectories and emerge from the solar surface. As they traverse the solar magnetic field, a fraction convert into photons, producing x-rays. The differential solar x-ray flux resulting from axion--photon conversion in the solar atmosphere is given by
\begin{equation}
    \frac{d\phi_\gamma}{dE_\gamma} = \frac{d\phi_a}{dE_a} \times P_{a\gamma},
    \label{eq:xray_flux}
\end{equation}
where $d\phi_a/dE_a$ is the total axion flux at the solar surface, obtained as the sum of the Primakoff, Bremsstrahlung, and Compton contributions in Eq.~\eqref{eq:solar_axion_flux_PBC}, and $P_{a\gamma}$ is the axion--photon conversion probability in the solar magnetic field.

To calculate the conversion probability, we start with the axion-photon mixing equation in an external magnetic field~\cite{Raffelt:1987im}
\begin{equation}
	\bigg( (E_a -i\partial_\ell)\mathbb{I} + M(\ell)\bigg)
	\begin{pmatrix}
		|A_\perp(\ell)\rangle\\
		|A_\parallel(\ell)\rangle\\
		|a(\ell)\rangle\\
	\end{pmatrix} = 0\,,
	\label{eq:EqofMotion}
\end{equation}
where $|A_{\parallel(\perp)}\rangle$ represents an electromagnetic wave polarized in the direction parallel (perpendicular) to $\vec{B}_T$ where $\vec{B}_T$ is the projection of the solar magnetic field $\vec{B}(\ell)$ on the plane perpendicular to the direction of propagation of the axions, and
\begin{equation}
	M(\ell) = 
	\begin{pmatrix}
		-\frac{\omega_p^2(\ell)}{2\omega} - i \dfrac{\Gamma(\ell)}{2} 
        &0 &0\\
		0 &-\frac{\omega_p^2(\ell)}{2\omega} - i \dfrac{\Gamma(\ell)}{2}         &g_{a\gamma\gamma}\frac{B_T(\ell)}{2}\\
		0 &g_{a\gamma\gamma}\frac{B_T(\ell)}{2} &-\frac{m_a^2}{2\omega}\\
	\end{pmatrix}\,.
\end{equation}
Here $\omega_p(\ell) = \sqrt{4\pi\, \alpha \,n_e(\ell)/m_e}$ is the plasma frequency in the solar atmosphere, with $\alpha$ being the fine structure constant and $\Gamma(\ell)$ is the coefficient of absorption of x-ray in the solar atmosphere~\cite{vanBibber:1988ge}. For this work, we neglect the Euler-Heisenberg term and Faraday rotation related to the longitudinal component of the external field. Since the axion field only couples to $A_\parallel$, axion-photon conversion can be described by the bottom $2\times2$ sub-block of $M$, which we denote as $\hat{M}$. Starting from the point $\ell = 0$, the solution of Eq. \eqref{eq:EqofMotion} at some location $\ell=h$ is given by the path-ordered transfer matrix~\cite{Raffelt:1987im,Manzari:2024jns}
\begin{align}
\label{eq:homsoln}
\begin{bmatrix}
    |A_\parallel(h)\rangle\\
    |a(h)\rangle
\end{bmatrix}
&=  {\cal P}_\ell\Bigg[
    \exp \Big( -i E_a h\, \mathbb{I}  -i \int_0^h \hat{M}(\ell) d \ell\Big)
\Bigg] \nonumber\\
&\quad \times
\begin{bmatrix}
    |A_\parallel(0)\rangle\\
    |a(0)\rangle
\end{bmatrix},
\end{align}
where $|A_\parallel(0)\rangle$ and $|a(0)\rangle$ are the initial field values.

The above integral can be evaluated perturbatively in powers of $\gagg$~\cite{Raffelt:1987im}. Working up to first order in perturbation theory, one can find that the probability of conversion of axions into photons after traveling a distance $h$ from the solar surface in the solar magnetic field as,
 \begin{align}
 \label{eq:osc_prob}
    P_{a\gamma}(h) =& \, |\langle A_\parallel (h) | a(0)\rangle|^2 =
    \dfrac{1}{4}\,\gagg^2\, e^{-\int_0^h\,d\ell\,\Gamma(\ell)}\,  \nonumber \\
   &\times \left| \int_0^h\, d\ell\, B_T(\ell)\, e^{i\int_0^\ell d\ell' q(\ell')}\,
   e^{\frac{1}{2} \int_0^\ell d\ell' \Gamma(\ell')}   \right|^2,
\end{align}
where $q(\ell) = (\omega_p^2 (\ell) - m_a^2)/2E_a$ is the momentum exchanged between the photon and the axion, both carrying energy $E_a$. 

\begin{figure}[ht!]
    \centering
    \includegraphics[width=0.82\linewidth]{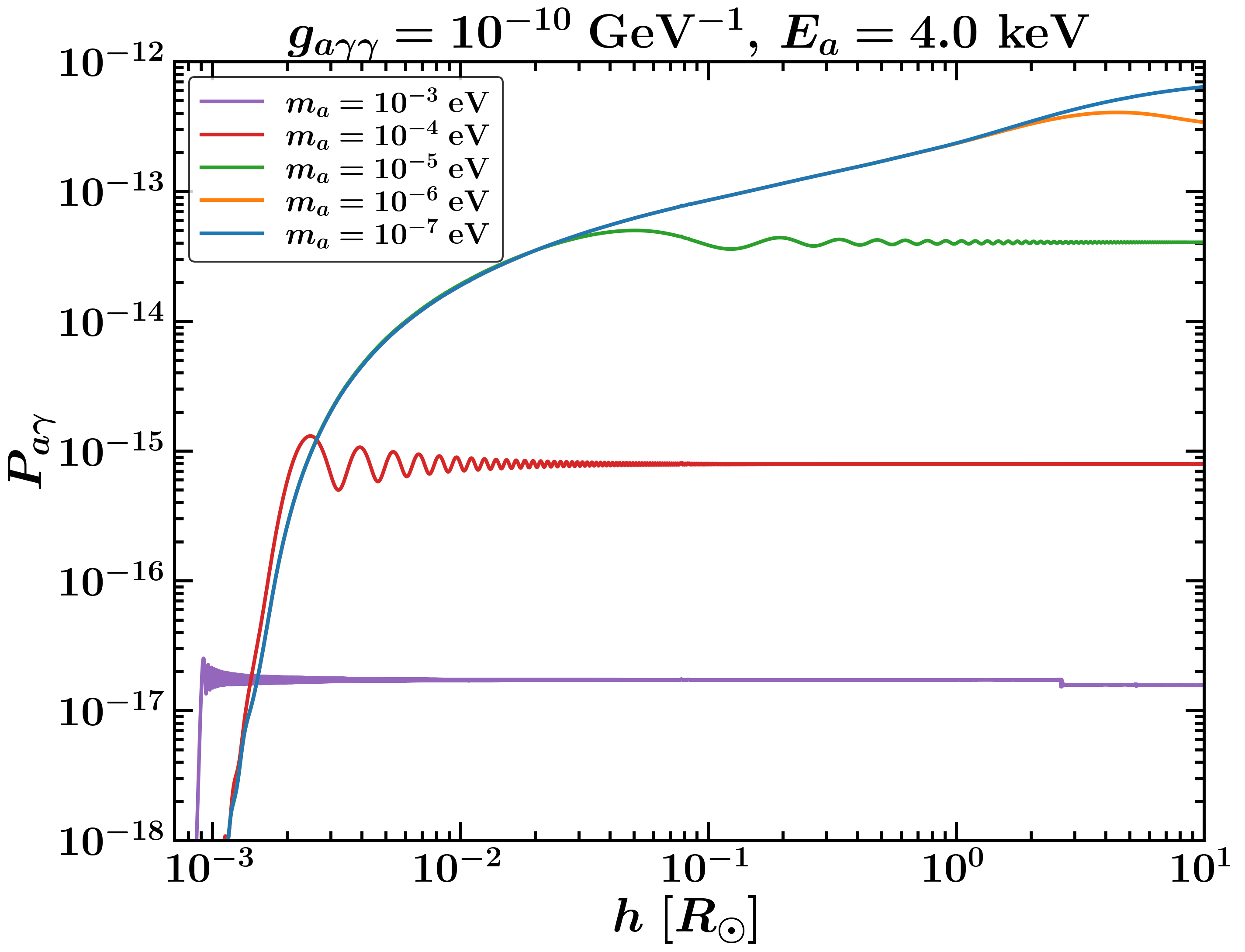}
    \includegraphics[width=0.82\linewidth]{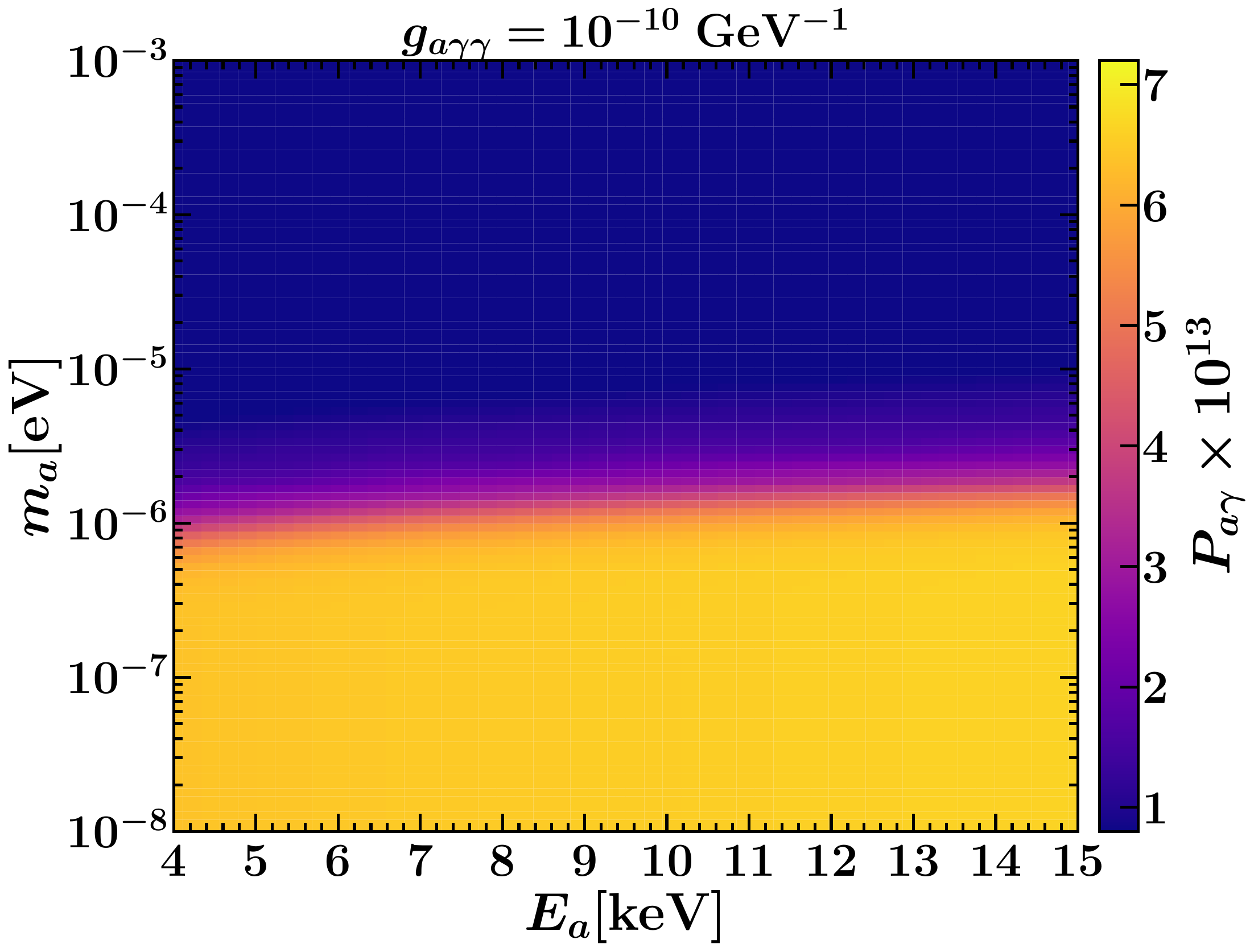}
    \caption{\textit{\textbf{Top panel:}} Variation of the axion to photon conversion probability $P_{a \gamma}$ with distance from the solar surface $h$ for fixed $\gagg$ and $E_a$ and for a range of $m_a$. \textit{\textbf{Bottom panel:}} Variation of $P_{a \gamma}$ with $m_a$ and $E_a$ for a fixed $\gagg$.}
    \label{fig:osc_prob}
\end{figure}

In this work, we compute the axion--photon conversion probability by numerically integrating Eq.~\eqref{eq:osc_prob} up to the required height $ h $ above the solar surface, using the magnetic-field and plasma-frequency profiles of the quiet Sun from Ref.~\cite{Ruz:2024gkl}. The absorption coefficient is computed as \textcolor{black}{$ \Gamma=\sum_{Z=1}^{30} n_Z \, \sigma_Z $}, where $n_Z$ is the number density of element of atomic number $Z$ and $\sigma_Z$ is the total attenuation cross-section  of a photon passing through a gas of atomic number $Z$. To compute $\Gamma$, we consider all elements up to atomic number $Z=30$. The individual elemental abundance in the solar atmosphere is obtained from the \texttt{CHIANTI} atomic database~\cite{2024ApJ...974...71D,2021ApJ...909...38D} by using the python interface \texttt{ChiantiPy}~\cite{ChiantiPy:2022xx}. For each element, we take the total photon attenuation cross-section with coherent scattering from the XCOM database by NIST~\cite{xcom}.

The top panel of Fig.~\ref{fig:osc_prob} shows the axion--photon conversion probability $ P_{a\gamma} $ as a function of height above the solar surface $h$ for various axion masses. The probability saturates at large distances and increases for lighter axions, becoming mass-independent for $ m_a \lesssim 10^{-7}\,\mathrm{eV} $, where the curves converge.

The bottom panel of Fig.~\ref{fig:osc_prob} shows $ P_{a\gamma} $ far from the Sun as a function of $ m_a $ and $ E_a $. The probability is almost independent of axion energy within the relevant range. Hence, from Eq.~\eqref{eq:xray_flux}, the resulting x-ray spectrum from axion conversion mirrors the solar axion spectrum in Fig.~\ref{fig:axion_flux}, differing only by an overall normalization set by the conversion probability.

\vspace*{0.5cm}
%
%
\begin{figure}[ht!]
    \centering
    \includegraphics[width=0.9\linewidth]{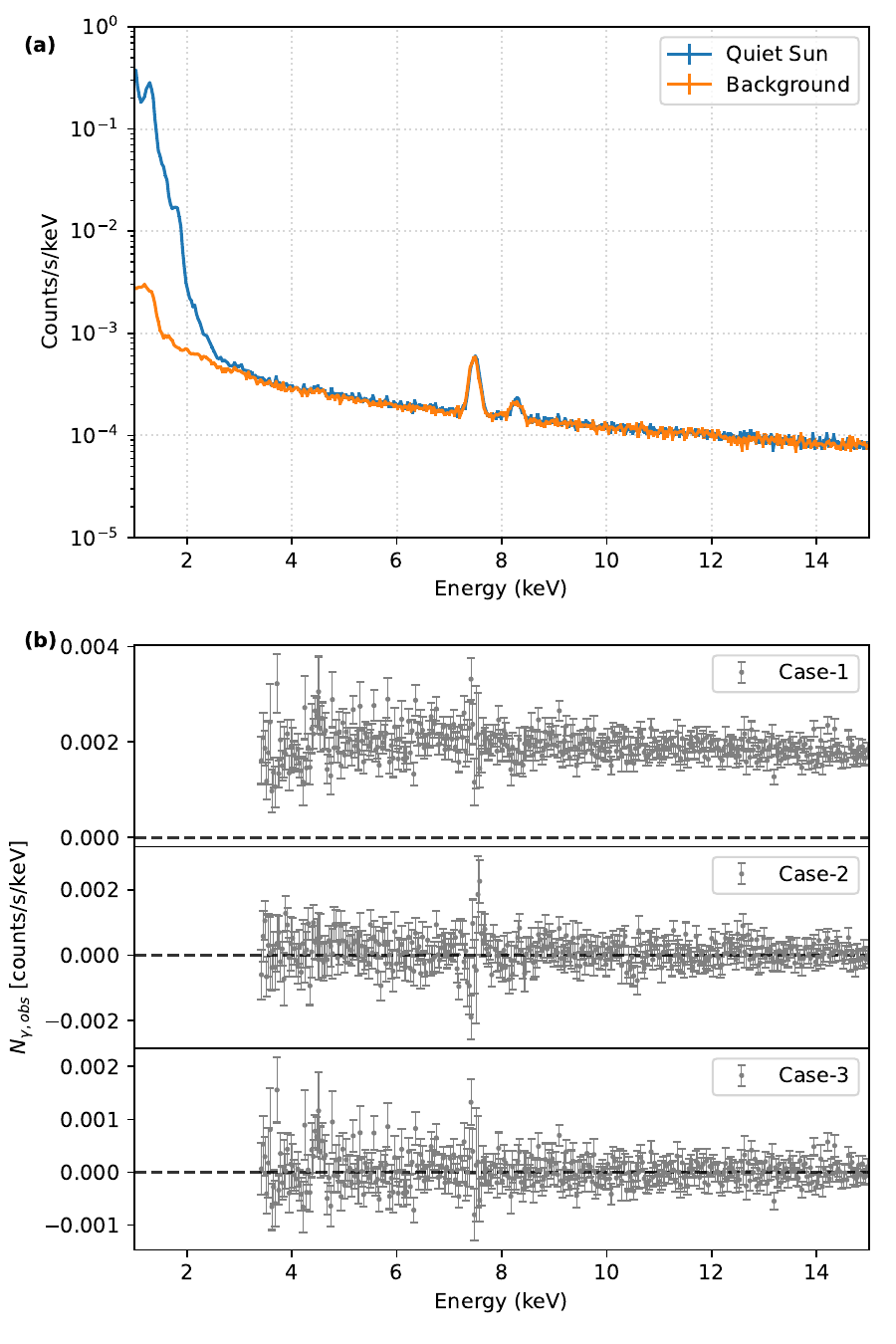}
        \caption{
        (\textbf{a}) XSM observed solar x-ray spectrum and background spectrum when the Sun is not in FOV. (\textbf{b}) XSM observed solar x-ray spectrum after background subtraction with three different methods \textit{\textbf{Top panel (Case 1):}} Only cosmic x-ray background subtraction along with avoiding bright x-ray source and energetic solar events (large SEP events), \textit{\textbf{Center panel (Case 2):}} Measured XSM background (when the Sun is out of FOV) subtraction  along with avoiding multiple bright x-ray sources and large SEP events, \textit{\textbf{Bottom panel (Case 3):}} Empirically modeled background subtraction.
        }
    \label{fig:xsm_data}
\end{figure}
\textit{Chandrayaan-2 observation of the quiet Sun.} Solar X-ray Monitor (XSM) on-board Chandrayaan-2 measures the disk-integrated solar x-ray spectrum in
1 -- 15 keV energy range~\cite{2020CSci..118...45S,2020SoPh..295..139M}. Here we use the XSM observations
during its first two observing seasons from September 2019 to May 2020, when the solar activity was at its minimum.
Data taken during the same observing periods of the quiet Sun when no active regions or microflares are present, as reported in
\citep{2021ApJ...912L..12V}, are used for the analysis to obtain the total quiet Sun x-ray spectrum.

In addition to emission from the Sun, XSM also records background which primarily consists of Cosmic X-ray Background (CXB),
charged particle background, and emission from other bright astrophysical sources. This background component needs to
be subtracted from the observations to obtain residual x-ray emission from the Sun. While the CXB component is steady,         
the measured CXB by XSM varies. Some part of the XSM FOV is usually obstructed by the Moon blocking the CXB emission,           
and the blocked fraction varies due to the variation in the orbital altitude of the spacecraft.
The charged particle background has temporal variations due to passage through geotail regions, solar energetic particle (SEP) events,
Coronal Mass Ejection (CME) events, which are also recorded by the upper level discriminator (ULD) counts of the XSM instrument~\cite{2020SoPh..295..139M}.
As the spacecraft attitude varies, bright astrophysical sources, Sco X-1 being the brightest source in this energy range,
enter within the XSM FOV for some periods and contribute to the background.
For the purpose of the present analysis, such periods of varying background with high ULD counts and presence of the Sco X-1 within the FOV are ignored 
while generating the quiet Sun x-ray spectrum to ensure that the background is as clean as possible.
We use the modules of XSM Data Analysis Software~\cite{2021A&C....3400449M} with custom good time intervals to generate
the quiet Sun spectrum, resulting in the net effective exposure of $3.72 \times \mathrm{10^6}$ s.

Fig.~\ref{fig:xsm_data}(a) shows the XSM quiet Sun spectrum in the 1--15 keV energy range, thus obtained
along with a background spectrum when the Sun is out of FOV.  The XSM spectrum of the quiet Sun is dominated
by the quiescent coronal emission in the low energies up to $\sim$3 keV\cite{2021ApJ...912L..12V} and
thus we use the spectrum above 3.4 keV to explore the x-ray emission produced by axion conversion.
To obtain the most conservative residual x-ray emission from the Sun, the CXB component alone is subtracted 
from this observed XSM spectrum. CXB model spectrum~\cite{2010A&A...512A..49T} taking into account the fraction 
of FOV obstructed by the Moon is subtracted from the observed XSM spectrum, and the resultant spectrum 
is shown in Fig.~\ref{fig:xsm_data}(b) top panel. The uncertainties shown are 1-sigma and arise from Poisson errors on the observed counts. The slight excess near 7.5 keV 
is due to a known instrumental line and may be ignored~\cite{2020SoPh..295..139M}.
As this spectrum includes the contribution from the charged particle background and other sources, a more realistic 
residual solar x-ray spectrum can be obtained by subtracting the background observed when the Sun is out of the FOV
from the quiet Sun observations. However, this requires ignoring the periods when other bright sources contribute
to the background. Sources that were bright enough to have a signal in XSM during the observing period were identified
from the  MAXI all-sky monitor data, and periods when such 13 sources were within the XSM FOV are ignored, resulting in an effective quiet Sun exposure of $2.15 \times \mathrm{10^6}$ s. The resulting spectrum after subtracting the background is shown in the middle panel of Fig.~\ref{fig:xsm_data}(b) (Case 2), errors incorporate uncertainties on the observed quiet Sun spectrum and the background spectrum.
We also consider Case 3, where the entire observed spectrum in Case 1 is assumed to arise from the background, and if the background were perfectly modeled, the resultant spectrum obtained is shown in the bottom panel of Fig.~\ref{fig:xsm_data}(b).
This is obtained by empirically modeling the observed spectrum and subtracting the best-fit model from the observed spectrum.

\vspace*{0.5cm}
\textit{Data analysis.}
The expected number of photons per unit time per unit energy in a given energy bin $i$ of the XSM solely due to axion-photon conversion in the solar atmosphere is
\begin{equation}
    N^i_{\gamma,\,\rm exp}  = \dfrac{R_\odot^2}{(\text{1 A.U.})^2}\,A_{\rm eff} (E_i)\,\dfrac{d \phi_a}{d E_a}(E_i)\,P_{a \gamma} (E_i),
\end{equation}
where $E_i$ is the energy of the $i$-th energy bin, \textcolor{black}{$A_{\rm eff}(E_i)$ is the effective collecting area of the XSM in the $i$-th energy bin},
and $R_\odot$ is the radius of the Sun. Given the observed number of counts per unit time per unit energy by the XSM, $N_{\gamma,\,obs}^i$, for a particular background subtraction scheme, we compute the likelihood assuming independent Gaussian statistic in each energy bin:
\begin{equation}
\mathcal{L}(g_{a\gamma\gamma}; m_a) \propto \prod_{i} \exp \left[ -\tfrac{1}{2} \left( \frac{N_{\gamma,\exp}^i - N_{\gamma,\text{obs}}^i}{\sigma^i} \right)^2 \right]
\end{equation}
where $\sigma^i$ denotes the uncertainty in the observed counts in the $i$-th energy bin. To compute our limits we consider all the energy bins between $3.4$ keV and $15$ keV.

To derive constraints in the \(m_a - \gagg\) plane, we fix the axion--electron coupling to \(g_{aee}=1.3\times10^{-13}\). For a given axion mass \(m_a\), we adopt a uniform prior on the axion--photon coupling \(\gagg\in[10^{-14},\,10^{-8}]~\mathrm{GeV}^{-1}\). The posterior probability density \(\mathcal{P}(\gagg\,|\,\text{data})\) is obtained within a Bayesian framework using the \texttt{PyMultiNest} implementation of the \texttt{MultiNest} algorithm~\cite{Feroz:2008xx}. The 95\% credible upper limit on \(\gagg\) is defined as the value enclosing 95\% of the posterior probability. Repeating this procedure for axion masses in the range \(m_a\in[10^{-12},\,10^{-3}]~\mathrm{eV}\) yields the corresponding mass-dependent upper limits on \(\gagg\).

\textcolor{black}{In addition, we also derive constraints in the \(g_{aee} - \,\gagg\) plane by fixing the axion mass to \(m_a=10^{-6}~\mathrm{eV}\). For \(m_a\lesssim10^{-4}~\mathrm{eV}\), the expected signal is effectively independent of \(m_a\), so this choice is representative of the entire low-mass regime. The analysis is repeated while scanning over \(g_{aee}\), yielding the corresponding 95\% credible upper limits on \(\gagg\) as a function of \(g_{aee}\).}

\begin{figure}
    \centering
    \includegraphics[width=0.82\linewidth]{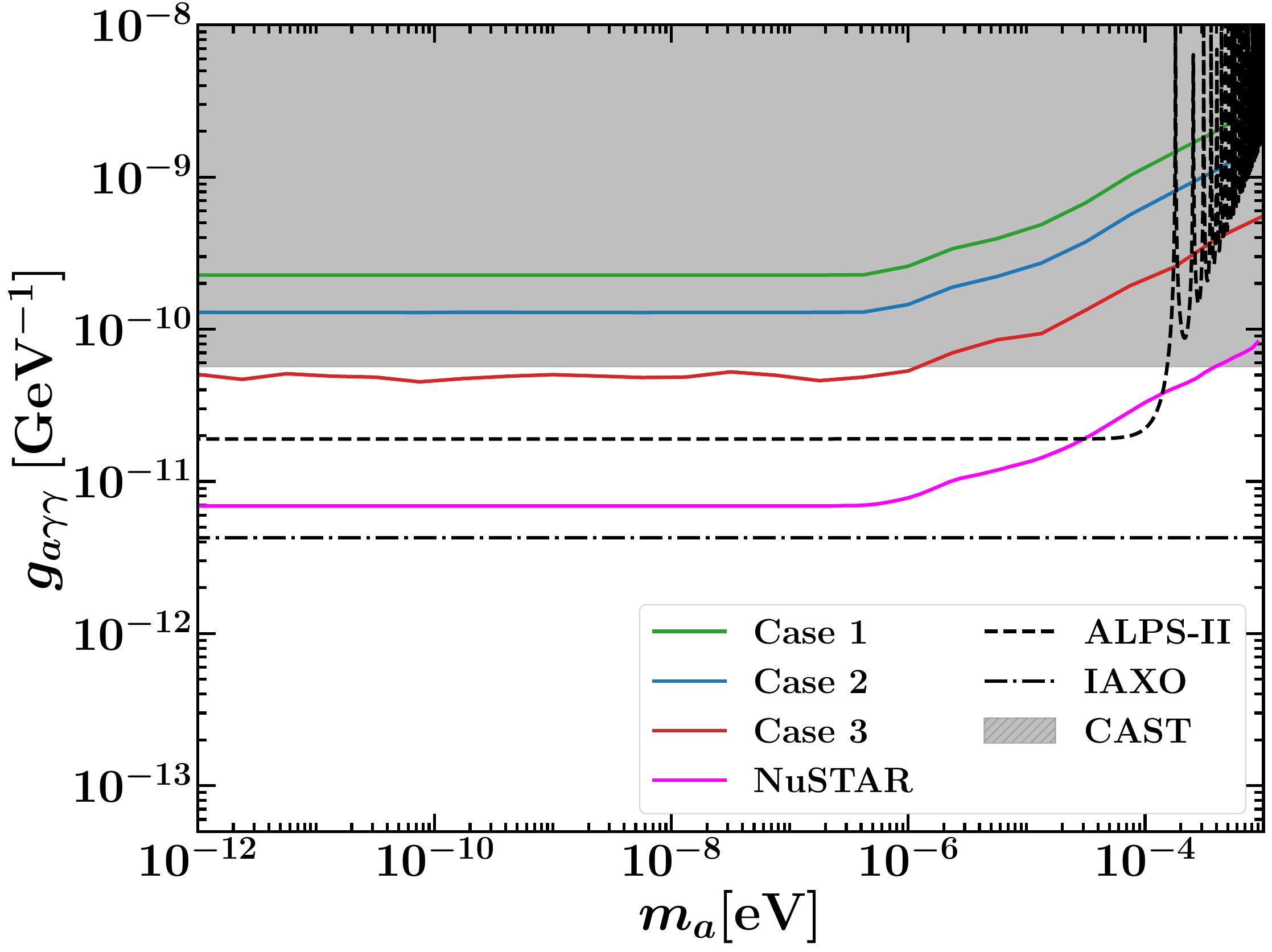}
    \includegraphics[width=0.82\linewidth]{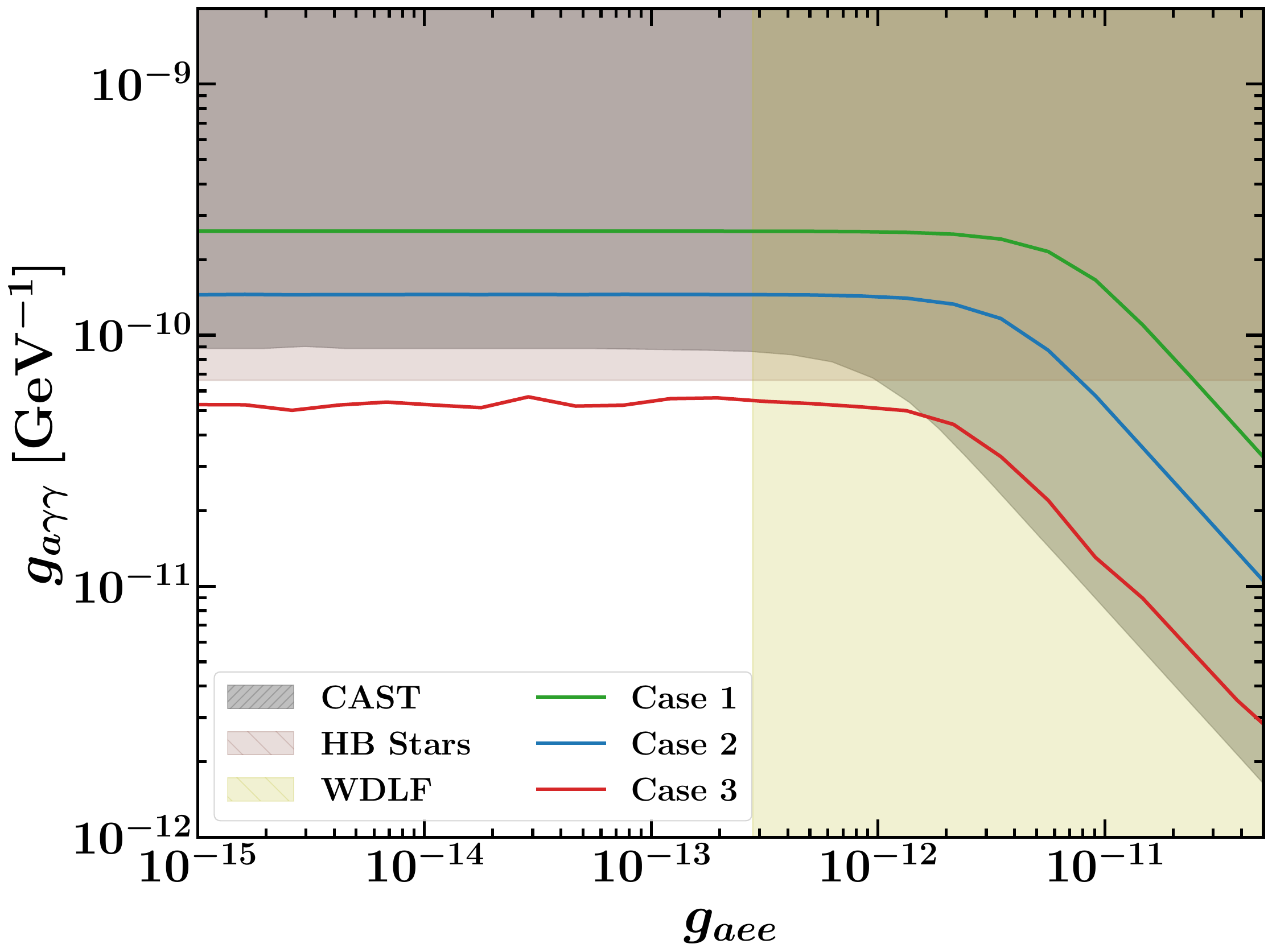}
    \caption{\textcolor{black}{Shown are the 95\% confidence level upper limits on $\gagg$ derived from the three background-subtracted solar X-ray spectra in Fig.~\ref{fig:xsm_data}. \textit{\textbf{Top Panel:}} Constraints in the $m_a –\, \gagg$ plane for a fixed $g_{ae}=1.3\times10^{-13}$, \textit{\textbf{Bottom Panel:}} Constraints in the $g_{aee} - \gagg$ plane obtained for a fixed $m_a=10^{-6}\,\mathrm{eV}$. Existing bounds and projected sensitivities are also shown for comparison.}}
    \label{fig:constraint_plot}
\end{figure}

\vspace*{0.5cm}
\textit{Results and discussion.} With the analysis described above, the observation of soft solar x-ray from quiet-Sun by the XSM onboard Chandrayaan-2 yields the following 95\% credible upper limit on the axion--photon coupling:
\begin{equation}
\textcolor{black}{g_{a\gamma\gamma} \lesssim (0.50\text{--}2.26)\times10^{-10}\,\mathrm{GeV}^{-1}},
\end{equation}
for axion masses $m_a \lesssim 5\times10^{-4}\,\mathrm{eV}$. Here the numerical spread reflects the variation due to the three background subtraction schemes adopted in this analysis: Case~1 (green solid line) gives \textcolor{black}{$g_{a\gamma\gamma}\lesssim 2.26\times10^{-10}\,\mathrm{GeV}^{-1}$} and is the most conservative limit, Case~2 (blue solid line) with realistic background subtraction gives \textcolor{black}{$g_{a\gamma\gamma}\lesssim 1.29\times10^{-10}\,\mathrm{GeV}^{-1}$}, and Case~3 (red solid line) yields the most stringent value \textcolor{black}{$g_{a\gamma\gamma}\lesssim 5.0\times10^{-11}\,\mathrm{GeV}^{-1}$} and is the target/optimistic limit from our analysis. These limits are shown in \textcolor{black}{the top panel of} Fig.~\ref{fig:constraint_plot}. The limits are effectively mass-independent across this low-mass range; sensitivity degrades for larger $m_a$ as coherence in the axion--photon conversion is lost when the axion mass exceeds the plasma frequency at the relevant solar-atmosphere heights.

To facilitate comparison with current and future laboratory searches, in Fig.~\ref{fig:constraint_plot} we overlay existing experimental limits from CAST~\cite{CAST:2017uph} (gray shaded region) and projected limits for ALPS-II~\cite{Bahre:2013ywa} (black dashed line) and IAXO~\cite{Armengaud:2014gea,IAXO:2019mpb} (black dot-dashed line). We also include the recent bound from NuSTAR~\cite{Ruz:2024gkl} (magenta line), which is currently the most stringent astrophysical constraint. While the most optimistic XSM limit (Case~3) remains about an order of magnitude weaker than the NuSTAR bound of $7.3\times10^{-12},\mathrm{GeV}^{-1}$, the XSM constraints are competitive with CAST and complementary to NuSTAR: XSM observes the full solar disk simultaneously, unlike NuSTAR, thereby avoiding systematics associated with partial-disk observations\footnote{The smaller effective collecting area of XSM largely accounts for the difference relative to NuSTAR ($A_{\rm eff}\simeq800,\mathrm{cm}^2$).}.

Complementary to the mass-dependent limits above, the bottom panel of Fig.~\ref{fig:constraint_plot} shows constraints in the $g_{aee}$--$\gagg$ plane, obtained by fixing \(m_a=10^{-6}\,\mathrm{eV}\), representative of the low-mass regime. The upper limits on \(\gagg\) strengthen monotonically with increasing $g_{aee}$, reflecting enhanced axion production in the solar interior. Existing bounds from horizontal-branch stars (brown shaded region), the white-dwarf luminosity function (olive shaded region), and CAST (gray shaded region) are shown for comparison~\cite{Barth:2013sma,Dent:2020jhf} and within the allowed parameter space, the Chandrayaan-2 XSM results do not significantly improve existing constraints on $g_{aee}$.

We have presented all three background treatments for two reasons. Firstly, showing Case~1 -- Case~3 makes explicit the degree to which the final sensitivity depends on background modeling; the spread quantifies the systematic uncertainty associated with background subtraction. Secondly, including Case~3 (the red/target curve) indicates the best sensitivity one may realistically expect from this dataset limited only by counting statistics, with an ideal background subtraction. 
The ideal background subtraction requires a comprehensive understanding of the various factors contributing to the detector background, including the cosmic X-ray background (CXB), the charged particle background of cosmic and solar origin, and secondary emission resulting from their interaction with the spacecraft structure. Apart from this, XSM, being a wide FOV instrument, is also affected by the variable bright astrophysical X-ray sources as well as the variable occultation of the CXB and particle background within the FOV by the Moon due to the altitude variation of the spacecraft. 
While Case~2 covers most of these aspects, some details, such as secondary emission from spacecraft structures and the time variability of all other X-ray sources within the FOV, which are very difficult to model, are not fully accounted for, and these will be necessary to reach the ideal Case~3.
In practice, Case~2 should be regarded as our baseline (realistic) constraint, Case~1 as the conservative bound, and Case~3 as the target limit that motivates further work on accurate background modeling.

Theoretical uncertainties in the solar axion flux are at the few-percent level and do not dominate the error budget. The consistency of the limits across the three background treatments—agreeing within an $\mathcal{O}(1)$ factor—shows that instrumental and modeling systematics are subdominant to statistical limitations, set primarily by the small XSM effective collecting area. The resulting constraints are also consistent with astrophysical requirements from solar neutrino measurements and helioseismology, which require axion-induced energy loss to remain a small fraction of the solar luminosity.

Finally, this remains an ongoing experiment. The comparison of the three limits demonstrates current robustness and defines clear targets for future improvements: increased exposure, improved background modeling, and coordinated solar monitoring can push sensitivity toward the Case~3 benchmark. Looking ahead, a next-generation solar x-ray observatory—especially near solar minimum—with a substantially larger collecting area and improved spectral resolution could achieve order-of-magnitude gains, probing parameter space complementary to planned helioscopes. The Chandrayaan-2 XSM results thus motivate dedicated space-based solar observatories for searches for axions and other physics beyond the Standard Model.

\vspace*{0.5cm}
\textit{Acknowledgements.}
TK would like to thank Heerak Banerjee and Elisa M. Todarello for helpful insights regarding the numerical caluclations and data analysis parts. Additionally TK would like to thank Sougata Ganguly and Rohan Pramanick for helpful discussions and comments. The authors acknowledge the use of data from the Solar X-ray Monitor (XSM) on board
the Chandrayaan-2 mission of the Indian Space Research Organisation
(ISRO), archived at the Indian Space Science Data Centre (ISSDC). XSM was
developed by Physical Research Laboratory (PRL) with support from various
ISRO centers. TK also acknowledges support in the form of Senior Research Fellowship from the Council of Scientific \& Industrial Research (CSIR), Government of India.

\bibliography{refs}

@ARTICLE{2021ApJ...912L..12V,
       author = "Vadawale, Santosh V. and others",
        title = "{Observations of the Quiet Sun during the Deepest Solar Minimum of the Past Century with Chandrayaan-2 XSM: Elemental Abundances in the Quiescent Corona}",
      journal = {Astrophys. J. Lett.},
         year = 2021,
        month = may,
       volume = {912},
       number = {1},
          eid = {L12},
        pages = {L12},
          doi = {10.3847/2041-8213/abf35d},
archivePrefix = {arXiv},
       eprint = {2103.16643},
 primaryClass = {astro-ph.SR},
       adsurl = {https://ui.adsabs.harvard.edu/abs/2021ApJ...912L..12V}
}

@ARTICLE{2020SoPh..295..139M,
       author = {{Mithun}, N.~P.~S. and others},
        title = "{Solar X-Ray Monitor on Board the Chandrayaan-2 Orbiter: In-Flight Performance and Science Prospects}",
      journal = {Sol. Phys.},
         year = 2020,
        month = oct,
       volume = {295},
       number = {10},
          eid = {139},
        pages = {139},
          doi = {10.1007/s11207-020-01712-1},
archivePrefix = {arXiv},
       eprint = {2009.09759},
 primaryClass = {astro-ph.SR},
       adsurl = {https://ui.adsabs.harvard.edu/abs/2020SoPh..295..139M}
}

@ARTICLE{2020CSci..118...45S,
       author = {{Shanmugam}, M. and and others},
        title = "{Solar X-ray Monitor onboard Chandrayaan-2 Orbiter}",
      journal = {Current Science},
         year = 2020,
        month = jan,
       volume = {118},
       number = {1},
        pages = {45},
          doi = {10.48550/arXiv.1910.09231},
archivePrefix = {arXiv},
       eprint = {1910.09231},
 primaryClass = {astro-ph.IM},
       adsurl = {https://ui.adsabs.harvard.edu/abs/2020CSci..118...45S}
}

@ARTICLE{2021A&C....3400449M,
       author = {{Mithun}, N.~P.~S. and and others},
        title = "{Data processing software for Chandrayaan-2 Solar X-ray Monitor}",
      journal = {Astronomy and Computing},
         year = 2021,
        month = jan,
       volume = {34},
          eid = {100449},
        pages = {100449},
          doi = {10.1016/j.ascom.2021.100449},
archivePrefix = {arXiv},
       eprint = {2007.11371},
 primaryClass = {astro-ph.IM},
       adsurl = {https://ui.adsabs.harvard.edu/abs/2021A&C....3400449M}
}

@ARTICLE{2010A&A...512A..49T,
       author = {{T{\"u}rler}, M. and and others},
        title = "{INTEGRAL hard X-ray spectra of the cosmic X-ray background and Galactic ridge emission}",
      journal = {Astronomy and Astrophysics},
         year = 2010,
        month = mar,
       volume = {512},
          eid = {A49},
        pages = {A49},
          doi = {10.1051/0004-6361/200913072},
archivePrefix = {arXiv},
       eprint = {1001.2110},
 primaryClass = {astro-ph.CO},
       adsurl = {https://ui.adsabs.harvard.edu/abs/2010A&A...512A..49T}
}

@article{Kim:2008hd,
    author = "Kim, Jihn E. and Carosi, Gianpaolo",
    title = "{Axions and the Strong CP Problem}",
    eprint = "0807.3125",
    archivePrefix = "arXiv",
    primaryClass = "hep-ph",
    doi = "10.1103/RevModPhys.82.557",
    journal = "Rev. Mod. Phys.",
    volume = "82",
    pages = "557--602",
    year = "2010",
    note = "[Erratum: Rev.Mod.Phys. 91, 049902 (2019)]"
}

@article{DiLuzio:2020wdo,
    author = "Di Luzio, Luca and others",
    title = "{The landscape of QCD axion models}",
    eprint = "2003.01100",
    archivePrefix = "arXiv",
    primaryClass = "hep-ph",
    reportNumber = "DESY 20-036, DESY-20-036",
    doi = "10.1016/j.physrep.2020.06.002",
    journal = "Phys. Rept.",
    volume = "870",
    pages = "1--117",
    year = "2020"
}

@article{Arvanitaki:2009fg,
    author = "Arvanitaki, Asimina and others",
    title = "{String Axiverse}",
    eprint = "0905.4720",
    archivePrefix = "arXiv",
    primaryClass = "hep-th",
    doi = "10.1103/PhysRevD.81.123530",
    journal = "Phys. Rev. D",
    volume = "81",
    pages = "123530",
    year = "2010"
}

@article{Barth:2013sma,
    author = "Barth, K. and others",
    title = "{CAST constraints on the axion-electron coupling}",
    eprint = "1302.6283",
    archivePrefix = "arXiv",
    primaryClass = "astro-ph.SR",
    doi = "10.1088/1475-7516/2013/05/010",
    journal = "JCAP",
    volume = "05",
    pages = "010",
    year = "2013"
}

@article{Raffelt:1985nk,
    author = "Raffelt, Georg G.",
    title = "{Astrophysical Axion Bounds Diminished By Screening Effects}",
    reportNumber = "MPI-PAE-PTH-51-85",
    doi = "10.1103/PhysRevD.33.897",
    journal = "Phys. Rev. D",
    volume = "33",
    pages = "897",
    year = "1986"
}

@article{Bahcall:2004pz,
    author = "Bahcall, John N. and others",
    title = "{New solar opacities, abundances, helioseismology, and neutrino fluxes}",
    eprint = "astro-ph/0412440",
    archivePrefix = "arXiv",
    doi = "10.1086/428929",
    journal = "Astrophys. J. Lett.",
    volume = "621",
    pages = "L85--L88",
    year = "2005"
}

@article{Peccei:1977hh,
    author = "Peccei, R. D. and Quinn, Helen R.",
    title = "{CP Conservation in the Presence of Instantons}",
    reportNumber = "ITP-568-STANFORD",
    doi = "10.1103/PhysRevLett.38.1440",
    journal = "Phys. Rev. Lett.",
    volume = "38",
    pages = "1440--1443",
    year = "1977"
}

@article{Peccei:1977ur,
    author = "Peccei, R. D. and Quinn, Helen R.",
    title = "{Constraints Imposed by CP Conservation in the Presence of Instantons}",
    reportNumber = "ITP-572-STANFORD",
    doi = "10.1103/PhysRevD.16.1791",
    journal = "Phys. Rev. D",
    volume = "16",
    pages = "1791--1797",
    year = "1977"
}

@article{Peccei:2006as,
    author = "Peccei, R. D.",
    editor = "Kuster, Markus and Raffelt, Georg and Beltran, Berta",
    title = "{The Strong CP problem and axions}",
    eprint = "hep-ph/0607268",
    archivePrefix = "arXiv",
    doi = "10.1007/978-3-540-73518-2_1",
    journal = "Lect. Notes Phys.",
    volume = "741",
    pages = "3--17",
    year = "2008"
}

@article{Weinberg:1977ma,
    author = "Weinberg, Steven",
    title = "{A New Light Boson?}",
    reportNumber = "HUTP-77/A074",
    doi = "10.1103/PhysRevLett.40.223",
    journal = "Phys. Rev. Lett.",
    volume = "40",
    pages = "223--226",
    year = "1978"
}

@article{Wilczek:1977pj,
    author = "Wilczek, Frank",
    title = "{Problem of Strong  $P$  and  $T$  Invariance in the Presence of Instantons}",
    reportNumber = "Print-77-0939 (COLUMBIA)",
    doi = "10.1103/PhysRevLett.40.279",
    journal = "Phys. Rev. Lett.",
    volume = "40",
    pages = "279--282",
    year = "1978"
}

@article{Raffelt:1987im,
    author = "Raffelt, Georg and Stodolsky, Leo",
    title = "{Mixing of the Photon with Low Mass Particles}",
    reportNumber = "MPI-PAE/PTh-54/87",
    doi = "10.1103/PhysRevD.37.1237",
    journal = "Phys. Rev. D",
    volume = "37",
    pages = "1237",
    year = "1988"
}

@article{Ouellet:2018beu,
    author = "Ouellet, Jonathan L. and others",
    title = "{First Results from ABRACADABRA-10 cm: A Search for Sub-$\mu$eV Axion Dark Matter}",
    eprint = "1810.12257",
    archivePrefix = "arXiv",
    primaryClass = "hep-ex",
    doi = "10.1103/PhysRevLett.122.121802",
    journal = "Phys. Rev. Lett.",
    volume = "122",
    number = "12",
    pages = "121802",
    year = "2019"
}

@article{Ouellet:2019tlz,
    author = "Ouellet, Jonathan L. and others",
    title = "{Design and implementation of the ABRACADABRA-10 cm axion dark matter search}",
    eprint = "1901.10652",
    archivePrefix = "arXiv",
    primaryClass = "physics.ins-det",
    doi = "10.1103/PhysRevD.99.052012",
    journal = "Phys. Rev. D",
    volume = "99",
    number = "5",
    pages = "052012",
    year = "2019"
}

@article{Silva-Feaver:2016qhh,
    author = "Silva-Feaver, Maximiliano and others",
    title = "{Design Overview of DM Radio Pathfinder Experiment}",
    eprint = "1610.09344",
    archivePrefix = "arXiv",
    primaryClass = "astro-ph.IM",
    doi = "10.1109/TASC.2016.2631425",
    journal = "IEEE Trans. Appl. Supercond.",
    volume = "27",
    number = "4",
    pages = "1400204",
    year = "2017"
}

@article{DMRadio:2022pkf,
    author = "Brouwer, L. and others",
    collaboration = "DMRadio",
    title = "{Projected sensitivity of DMRadio-m3: A search for the QCD axion below 1\,\,\ensuremath{\mu}eV}",
    eprint = "2204.13781",
    archivePrefix = "arXiv",
    primaryClass = "hep-ex",
    doi = "10.1103/PhysRevD.106.103008",
    journal = "Phys. Rev. D",
    volume = "106",
    number = "10",
    pages = "103008",
    year = "2022"
}

@article{DMRadio:2022jfv,
    author = "Brouwer, L. and others",
    collaboration = "DMRadio",
    title = "{Proposal for a definitive search for GUT-scale QCD axions}",
    eprint = "2203.11246",
    archivePrefix = "arXiv",
    primaryClass = "hep-ex",
    doi = "10.1103/PhysRevD.106.112003",
    journal = "Phys. Rev. D",
    volume = "106",
    number = "11",
    pages = "112003",
    year = "2022"
}

@article{ADMX:2003rdr,
    author = "Asztalos, S. J. and others",
    collaboration = "ADMX",
    title = "{An Improved RF cavity search for halo axions}",
    eprint = "astro-ph/0310042",
    archivePrefix = "arXiv",
    doi = "10.1103/PhysRevD.69.011101",
    journal = "Phys. Rev. D",
    volume = "69",
    pages = "011101",
    year = "2004"
}

@article{ADMX:2018gho,
    author = "Du, N. and others",
    collaboration = "ADMX",
    title = "{A Search for Invisible Axion Dark Matter with the Axion Dark Matter Experiment}",
    eprint = "1804.05750",
    archivePrefix = "arXiv",
    primaryClass = "hep-ex",
    reportNumber = "FERMILAB-PUB-18-101-AD-AE",
    doi = "10.1103/PhysRevLett.120.151301",
    journal = "Phys. Rev. Lett.",
    volume = "120",
    number = "15",
    pages = "151301",
    year = "2018"
}

@article{ADMX:2019uok,
    author = "Braine, T. and others",
    collaboration = "ADMX",
    title = "{Extended Search for the Invisible Axion with the Axion Dark Matter Experiment}",
    eprint = "1910.08638",
    archivePrefix = "arXiv",
    primaryClass = "hep-ex",
    reportNumber = "FERMILAB-PUB-19-569-AD-AE-PPD",
    doi = "10.1103/PhysRevLett.124.101303",
    journal = "Phys. Rev. Lett.",
    volume = "124",
    number = "10",
    pages = "101303",
    year = "2020"
}

@article{HAYSTAC:2018rwy,
    author = "Zhong, L. and others",
    collaboration = "HAYSTAC",
    title = "{Results from phase 1 of the HAYSTAC microwave cavity axion experiment}",
    eprint = "1803.03690",
    archivePrefix = "arXiv",
    primaryClass = "hep-ex",
    doi = "10.1103/PhysRevD.97.092001",
    journal = "Phys. Rev. D",
    volume = "97",
    number = "9",
    pages = "092001",
    year = "2018"
}

@article{Caldwell:2016dcw,
    author = "Caldwell, Allen and others",
    collaboration = "MADMAX Working Group",
    title = "{Dielectric Haloscopes: A New Way to Detect Axion Dark Matter}",
    eprint = "1611.05865",
    archivePrefix = "arXiv",
    primaryClass = "physics.ins-det",
    doi = "10.1103/PhysRevLett.118.091801",
    journal = "Phys. Rev. Lett.",
    volume = "118",
    number = "9",
    pages = "091801",
    year = "2017"
}

@article{Millar:2016cjp,
    author = "Millar, Alexander J. and others",
    title = "{Dielectric Haloscopes to Search for Axion Dark Matter: Theoretical Foundations}",
    eprint = "1612.07057",
    archivePrefix = "arXiv",
    primaryClass = "hep-ph",
    doi = "10.1088/1475-7516/2017/01/061",
    journal = "JCAP",
    volume = "01",
    pages = "061",
    year = "2017"
}

@article{CAST:2004gzq,
    author = "Zioutas, K. and others",
    collaboration = "CAST",
    title = "{First results from the CERN Axion Solar Telescope (CAST)}",
    eprint = "hep-ex/0411033",
    archivePrefix = "arXiv",
    doi = "10.1103/PhysRevLett.94.121301",
    journal = "Phys. Rev. Lett.",
    volume = "94",
    pages = "121301",
    year = "2005"
}

@article{CAST:2017uph,
    author = "Anastassopoulos, V. and others",
    collaboration = "CAST",
    title = "{New CAST Limit on the Axion-Photon Interaction}",
    eprint = "1705.02290",
    archivePrefix = "arXiv",
    primaryClass = "hep-ex",
    doi = "10.1038/nphys4109",
    journal = "Nature Phys.",
    volume = "13",
    pages = "584--590",
    year = "2017"
}

@article{Dent:2020jhf,
    author = "Dent, James B. and others",
    title = "{Inverse Primakoff Scattering as a Probe of Solar Axions at Liquid Xenon Direct Detection Experiments}",
    eprint = "2006.15118",
    archivePrefix = "arXiv",
    primaryClass = "hep-ph",
    reportNumber = "MI-TH-2018",
    doi = "10.1103/PhysRevLett.125.131805",
    journal = "Phys. Rev. Lett.",
    volume = "125",
    number = "13",
    pages = "131805",
    year = "2020"
}

@article{IAXO:2019mpb,
    author = "Armengaud, E. and others",
    collaboration = "IAXO",
    title = "{Physics potential of the International Axion Observatory (IAXO)}",
    eprint = "1904.09155",
    archivePrefix = "arXiv",
    primaryClass = "hep-ph",
    doi = "10.1088/1475-7516/2019/06/047",
    journal = "JCAP",
    volume = "06",
    pages = "047",
    year = "2019"
}

@article{Vadawale:2021pis,
    author = "Vadawale, Santosh V. and others",
    title = "{Observations of the Quiet Sun during the Deepest Solar Minimum of the Past Century with Chandrayaan-2 XSM: Sub-A-class Microflares outside Active Regions}",
    eprint = "2103.16644",
    archivePrefix = "arXiv",
    primaryClass = "astro-ph.SR",
    doi = "10.3847/2041-8213/abf0b0",
    journal = "Astrophys. J. Lett.",
    volume = "912",
    number = "1",
    pages = "L13",
    year = "2021"
}

@article{Ruz:2024gkl,
    author = "Ruz, J. and others",
    title = "{NuSTAR as an Axion Helioscope}",
    eprint = "2407.03828",
    archivePrefix = "arXiv",
    primaryClass = "astro-ph.CO",
    month = "7",
    journal = "",
    year = "2024"
}

@article{Manzari:2024jns,
    author = "Manzari, Claudio Andrea and others",
    title = "{Supernova Axions Convert to Gamma Rays in Magnetic Fields of Progenitor Stars}",
    eprint = "2405.19393",
    archivePrefix = "arXiv",
    primaryClass = "hep-ph",
    doi = "10.1103/PhysRevLett.133.211002",
    journal = "Phys. Rev. Lett.",
    volume = "133",
    number = "21",
    pages = "211002",
    year = "2024"
}

@ARTICLE{2024ApJ...974...71D,
       author = {{Dufresne}, R.~P. and others},
        title = "{CHIANTI{\textemdash}An Atomic Database for Emission Lines{\textemdash}Paper. XVIII. Version 11, Advanced Ionization Equilibrium Models: Density and Charge Transfer Effects}",
      journal = {\apj},
         year = 2024,
        month = oct,
       volume = {974},
       number = {1},
          eid = {71},
        pages = {71},
          doi = {10.3847/1538-4357/ad6765},
archivePrefix = {arXiv},
       eprint = {2403.16922},
 primaryClass = {astro-ph.SR},
       adsurl = {https://ui.adsabs.harvard.edu/abs/2024ApJ...974...71D}
}

@ARTICLE{2021ApJ...909...38D,
       author = {{Del Zanna}, G. and others},
        title = "{CHIANTI{\textemdash}An Atomic Database for Emission Lines. XVI. Version 10, Further Extensions}",
      journal = {\apj},
         year = 2021,
        month = mar,
       volume = {909},
       number = {1},
          eid = {38},
        pages = {38},
          doi = {10.3847/1538-4357/abd8ce},
archivePrefix = {arXiv},
       eprint = {2011.05211},
 primaryClass = {physics.atom-ph},
       adsurl = {https://ui.adsabs.harvard.edu/abs/2021ApJ...909...38D}
}

@misc{ChiantiPy:2022xx,
  title        = {ChiantiPy},
  year         = {2022},
  howpublished = {\url{https://github.com/chianti-atomic/ChiantiPy/}}
}

@online{xcom,
  author = {Berger, M. J. and others},
  title = {{XCOM}: {Photon Cross Section Database}},
  year         = {2010},
  url          = {https://www.nist.gov/pml/xcom-photon-cross-sections-database}
}

@article{Feroz:2008xx,
    author = "Feroz, F. and others",
    title = "{MultiNest: an efficient and robust Bayesian inference tool for cosmology and particle physics}",
    eprint = "0809.3437",
    archivePrefix = "arXiv",
    primaryClass = "astro-ph",
    doi = "10.1111/j.1365-2966.2009.14548.x",
    journal = "Mon. Not. Roy. Astron. Soc.",
    volume = "398",
    pages = "1601--1614",
    year = "2009"
}

@article{Bahre:2013ywa,
    author = {B{\"a}hre, Robin and others},
    title = "{Any light particle search II {\textemdash}Technical Design Report}",
    eprint = "1302.5647",
    archivePrefix = "arXiv",
    primaryClass = "physics.ins-det",
    reportNumber = "DESY-13-030",
    doi = "10.1088/1748-0221/8/09/T09001",
    journal = "JINST",
    volume = "8",
    pages = "T09001",
    year = "2013"
}

@article{Armengaud:2014gea,
    author = "Armengaud, E. and others",
    title = "{Conceptual Design of the International Axion Observatory (IAXO)}",
    eprint = "1401.3233",
    archivePrefix = "arXiv",
    primaryClass = "physics.ins-det",
    reportNumber = "FERMILAB-PUB-14-476-A, BNL-106253-2014-JA",
    doi = "10.1088/1748-0221/9/05/T05002",
    journal = "JINST",
    volume = "9",
    pages = "T05002",
    year = "2014"
}

@article{Preskill:1982cy,
    author = "Preskill, John and Wise, Mark B. and Wilczek, Frank",
    editor = "Srednicki, M. A.",
    title = "{Cosmology of the Invisible Axion}",
    reportNumber = "HUTP-82-A048, NSF-ITP-82-103",
    doi = "10.1016/0370-2693(83)90637-8",
    journal = "Phys. Lett. B",
    volume = "120",
    pages = "127--132",
    year = "1983"
}

@article{Abbott:1982af,
    author = "Abbott, L. F. and Sikivie, P.",
    editor = "Srednicki, M. A.",
    title = "{A Cosmological Bound on the Invisible Axion}",
    reportNumber = "PRINT-82-0695 (BRANDEIS)",
    doi = "10.1016/0370-2693(83)90638-X",
    journal = "Phys. Lett. B",
    volume = "120",
    pages = "133--136",
    year = "1983"
}

@article{Dine:1982ah,
    author = "Dine, Michael and Fischler, Willy",
    editor = "Srednicki, M. A.",
    title = "{The Not So Harmless Axion}",
    reportNumber = "UPR-0201T",
    doi = "10.1016/0370-2693(83)90639-1",
    journal = "Phys. Lett. B",
    volume = "120",
    pages = "137--141",
    year = "1983"
}

@article{Khlopov:1999tm,
    author = "Khlopov, M. Yu. and others",
    editor = "Sikivie, P.",
    title = "{The nonlinear modulation of the density distribution in standard axionic CDM and its cosmological impact}",
    doi = "10.1016/S0920-5632(98)00511-8",
    journal = "Nucl. Phys. B Proc. Suppl.",
    volume = "72",
    pages = "105--109",
    year = "1999"
}

@article{Vysotsky:1978dc,
    author = "Vysotsky, M. I. and others",
    title = "{Some Astrophysical Limitations on Axion Mass}",
    journal = "Pisma Zh. Eksp. Teor. Fiz.",
    volume = "27",
    pages = "533--536",
    year = "1978"
}

@article{Raffelt:2006cw,
    author = "Raffelt, Georg G.",
    editor = "Kuster, Markus and Raffelt, Georg and Beltran, Berta",
    title = "{Astrophysical axion bounds}",
    eprint = "hep-ph/0611350",
    archivePrefix = "arXiv",
    reportNumber = "MPP-2006-172",
    doi = "10.1007/978-3-540-73518-2_3",
    journal = "Lect. Notes Phys.",
    volume = "741",
    pages = "51--71",
    year = "2008"
}

@article{Lella:2022uwi,
    author = "Lella, Alessandro and others",
    title = "{Protoneutron stars as cosmic factories for massive axionlike particles}",
    eprint = "2211.13760",
    archivePrefix = "arXiv",
    primaryClass = "hep-ph",
    doi = "10.1103/PhysRevD.107.103017",
    journal = "Phys. Rev. D",
    volume = "107",
    number = "10",
    pages = "103017",
    year = "2023"
}

@article{Lella:2023bfb,
    author = "Lella, Alessandro and others",
    title = "{Getting the most on supernova axions}",
    eprint = "2306.01048",
    archivePrefix = "arXiv",
    primaryClass = "hep-ph",
    doi = "10.1103/PhysRevD.109.023001",
    journal = "Phys. Rev. D",
    volume = "109",
    number = "2",
    pages = "023001",
    year = "2024"
}

@article{Carenza:2024ehj,
    author = "Carenza, Pierluca and others",
    title = "{Axion astrophysics}",
    eprint = "2411.02492",
    archivePrefix = "arXiv",
    primaryClass = "hep-ph",
    reportNumber = "BARI-TH/66-24",
    doi = "10.1016/j.physrep.2025.02.002",
    journal = "Phys. Rept.",
    volume = "1117",
    pages = "1--102",
    year = "2025"
}

@article{Lecce:2025dbz,
    author = "Lecce, Francesca and others",
    title = "{Probing axionlike particles with multimessenger observations of neutron star mergers}",
    eprint = "2504.02032",
    archivePrefix = "arXiv",
    primaryClass = "hep-ph",
    reportNumber = "BARI-TH/773-25",
    doi = "10.1103/krf3-lm4s",
    journal = "Phys. Rev. D",
    volume = "112",
    number = "2",
    pages = "023001",
    year = "2025"
}

@article{Fiorillo:2025gnd,
    author = "Fiorillo, Damiano F. G. and others",
    title = "{Axion-photon conversion in transient compact stars: Systematics, constraints, and opportunities}",
    eprint = "2509.13322",
    archivePrefix = "arXiv",
    primaryClass = "hep-ph",
    month = "9",
    year = "2025",
journal = ""
}

@article{vanBibber:1988ge,
    author = "van Bibber, K. and others",
    title = "{A Practical Laboratory Detector for Solar Axions}",
    reportNumber = "LBL-25908",
    doi = "10.1103/PhysRevD.39.2089",
    journal = "Phys. Rev. D",
    volume = "39",
    pages = "2089",
    year = "1989"
}

\end{document}